\documentclass{elsart}

\usepackage{epsfig}
\begin{document}
\topmargin = 1.cm
\begin{frontmatter}
%
%
\title{CP Trajectory Diagram; \\
--- A tool for pictorial representation of CP and matter effects 
in neutrino oscillations ---}
\thanks[label1]{Talk presented at The 3rd International Workshop on 
Neutrino Factories Based on Muon Storage Rings (NuFACT01), 
Tsukuba, Japan, May 24-30, 2001.}
%
\author[label1]{Hisakazu Minakata} and  
\author[label2,label3]{Hiroshi Nunokawa}
\address[label1]{Department of Physics, Tokyo Metropolitan University \\
1-1 Minami-Osawa, Hachioji, Tokyo 192-0397, Japan}
\address[label2]
{Instituto de F\'{\i}sica Te\'orica, 
Universidade Estadual Paulista \\
Rua Pamplona 145, 01405-900 S\~ao Paulo, SP Brazil}
\address[label3]
{Instituto de F\'{\i}sica Gleb Wataghin, Universidade Estadual
de Campinas \\ 
P.O. Box 6165, 13083-970 Campinas SP Brazil
}
%
\begin{abstract}
We introduce ``CP trajectory diagram in bi-probability space" 
as a powerful tool for pictorial representation of the genuine CP 
and the matter effects in neutrino oscillations. Existence of the 
correlated ambiguity in a determination of CP violating phase 
$\delta$ and the sign of $\Delta m^2_{13}$ is uncovered.
Principles of tuning beam energy for a given baseline distance 
are proposed to resolve the ambiguity and to maximize the 
CP-odd effect. We finally point out, quite contrary to what 
is usually believed, that the ambiguity may be resolved with 
50 \% chance in the super-JHF experiment despite its relatively short 
baseline of 300 km. 

\end{abstract}
\end{frontmatter}

\section{Introduction}

Probing into CP violation in the lepton sector is one of the 
most challenging goals in particle physics. Long baseline neutrino 
oscillation experiments are the prime candidates for observational 
means for detecting such effect. 
However, it has been known since sometime ago that the earth matter 
effect acts as a contamination to the measurement of genuine 
CP violating effects due to the leptonic Kobayashi-Maskawa phase 
in such experiments \cite{cp-matter}. Therefore, 
it is of crucial importance to achieve a complete understanding 
of the features of interplay between the genuine CP and the matter 
effects to have a realistic design of experiments for measuring 
the CP violating phase. 

We describe in this article a new powerful tool which we call 
``CP trajectory diagram in bi-probability space" \cite {MNjhep01}. 
It enables us to represent pictorially the three effects, 
the effects of 
(a) genuine CP violation due to the $\sin \delta$ term, 
(b) CP conserving $\cos \delta$ term, and 
(c) fake CP violation due to earth matter, 
separately in a single diagram. 
By using the CP trajectory diagram we observe that there is a
two-fold ambiguity in the determination of $\delta$ which is
related with the sign of $\Delta m^2_{13}$. As described in 
detail in ref. \cite {MNjhep01}, this is a remnant of the approximate 
degeneracy in the vacuum oscillation probability under 
the transformations 
($\delta \rightarrow \pi - \delta$) and 
($\Delta m^2_{13} \rightarrow - \Delta m^2_{13}$).

We then discuss principles of tuning beam energy for a given 
baseline distance to resolve the ambiguity, and to maximize the 
CP-violating effect. Finally, we point out that the ambiguity 
may be resolved with 50 \% chance in the super-JHF experiment 
\cite {JHF} with a megaton class water Cherenkov detector. 
It is quite contrary to the conventional belief that the sign 
of $\Delta m^2_{13}$ can not be determined with such 
a short baseline as $L=300$ km.
In a companion article, which is a report for the Proceedings 
of TAUP2001 \cite{taup2001}, we discuss further physical 
implications of our results. In particular, we explore the 
possibility of an {\it in situ} simultaneous measurement 
of $\delta$ and the sign of $\Delta m^2_{13}$ in a single 
experiment.

\section{CP trajectory diagram in bi-probability space}

We now introduce the CP trajectory diagram
in bi-probability space spanned by 
$P(\nu) \equiv P(\nu_{\mu} \rightarrow \nu_{e})$ and 
$P(\bar{\nu}) \equiv P(\bar{\nu}_{\mu} \rightarrow \bar{\nu}_{e})$. 
Suppose that we compute the oscillation probability $P(\nu)$ and
$P(\bar{\nu})$ with a given set of oscillation and experimental
parameters. Then, we draw a dot on the two-dimensional plane spanned
by $P(\nu)$ and $P(\bar{\nu})$.
When $\delta$ is varied we have a set of dots which forms a closed
trajectory, closed because the probability must be a periodic
function of $\delta$, a phase variable.

In fig. 1 plotted is the contours of oscillation probabilities 
$P(\nu)$ and $P(\bar{\nu})$ which is drawn by varying the CP 
violating phase $\delta$ from 0 to $2\pi$.
They are averaged over the Gaussian neutrino 
energy distribution with peak energy of 500 MeV and width 100 MeV. 
As you might have guessed, these diagrams are elliptic. It is so 
exactly in vacuum and in a good approximation even in matter 
at relatively short baseline and as far as the mass hierarchy 
$|\Delta m^2_{12}/\Delta m^2_{13}| \ll 1$ is the case 
\cite {MNjhep01}.

What does CP trajectory diagram actually represent?
There is a very simple answer to this question. 
The lengths of major and minor axes represent the coefficients 
of CP violating $\sin{\delta}$ and CP conserving $\cos{\delta}$ 
terms, respectively, in the neutrino oscillation probability 
$P(\nu_{\mu} \rightarrow \nu_{e})$. Whereas the distance between 
two (positive and negative $\Delta m^2_{13}$) ellipses gives 
the size of the matter effect. 
The last point can be explicitly verified by running 
the same calculation with twice larger matter 
effects, as done in ref. \cite {MNjhep01}.
Therefore, you can see the size of these three terms just by eye 
when it was projected onto the CP trajectory diagram. 

\vglue 0.3cm
\begin{center}
\centerline{\psfig{file=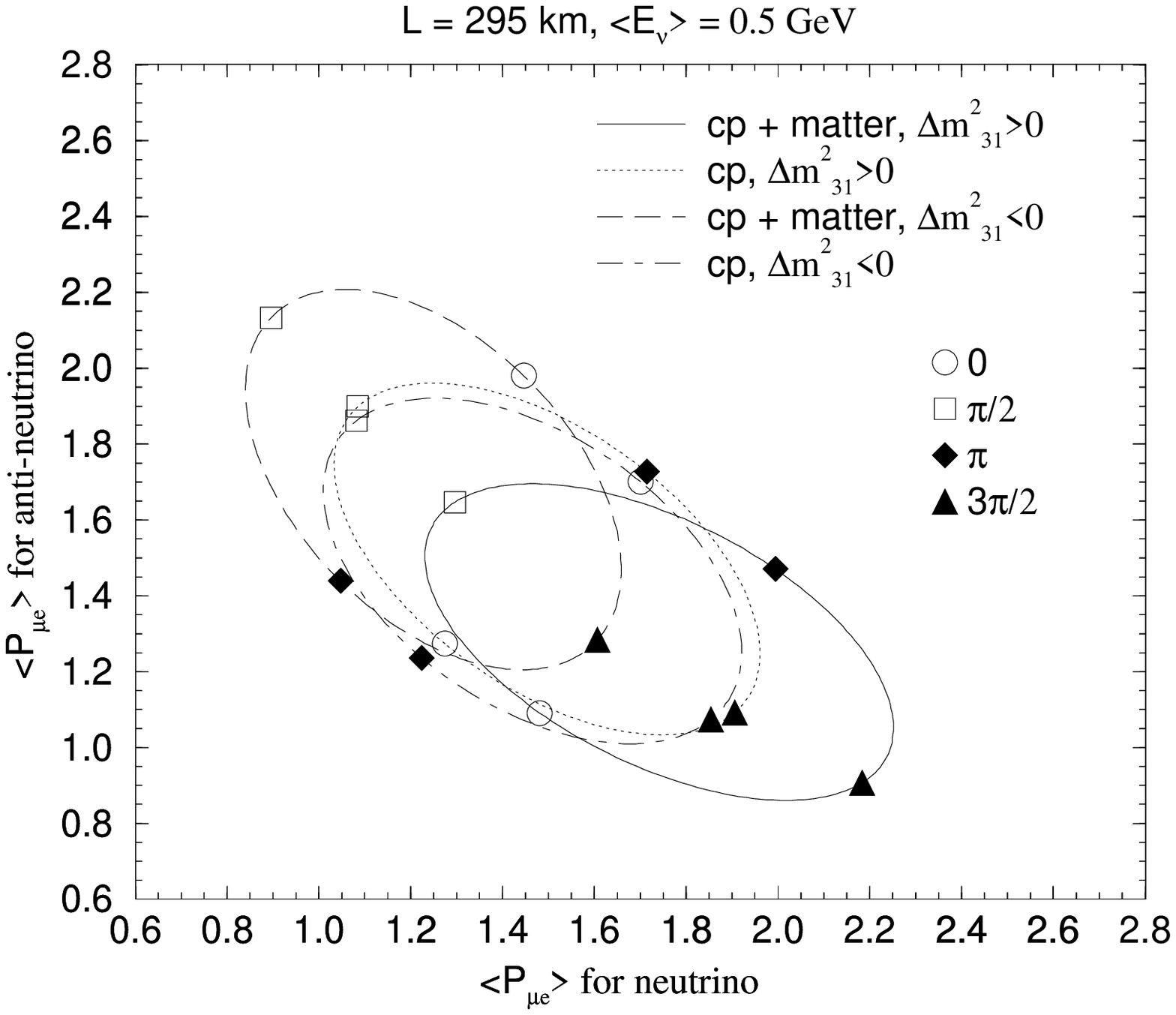,height=12cm,width=12cm}}
\end{center}
\vglue -2.0cm
\noindent
{Fig. 1: CP trajectory in the bi-probability (given in \%) plane
for the baseline $L=295$ km and energy 
$\langle E \rangle  =500$ MeV.
As indicated in the figures, the solid and the dashed lines are for
$\Delta m^2_{13} > 0$ and $\Delta m^2_{13} < 0$ cases, respectively,
and the dotted and the dash-dotted lines correspond to the same
signs of $\Delta m^2_{13}$ as above but with matter effect switched off.
The mixing parameters are fixed as
$\Delta m^2_{13}
 = \pm 3 \times 10^{-3}$ eV$^2$,
$\sin^22\theta_{23} = 1.0$,
$\Delta m^2_{12} = 5\times 10^{-5}$ eV$^2$,
$\sin^22\theta_{12} = 0.8$,
$\sin^22\theta_{13} = 0.05$.
We take $\rho Y_e = 1.4$ g/cm$^3$ where $\rho$ is
the matter density and $Y_e$ is the electron fraction.}
\vglue 0.5cm

It is obvious from fig. 1 that the approximate degeneracy 
($\delta \rightarrow \pi - \delta$) and 
($\Delta m^2_{13} \rightarrow - \Delta m^2_{13}$) which exists 
in the vacuum case is lifted by the matter effect; matter helps!
However, it is also clearly seen in fig. 1 that there 
are remaining degeneracies if the sign of $\Delta m^2_{13}$ 
is not known {\it a priori}.

The above discussion assumes that the value of $\theta_{13}$ is 
known prior to the experiment. If the value is not known in advance 
there is an another ambiguity, the ($\delta - \theta_{13}$) ambiguity, 
as discussed in detail in ref. \cite{B-Castell}. 
Therefore, there exits a combined ambiguity,  
($\delta$ $-$ sign of $\Delta m^2_{13}$) times 
($\delta - \theta_{13}$), which in the worst case can be 4-fold. 
The ambiguity was referred to as the ``clover-leaf ambiguity'' 
in TAUP2001 \cite{taup2001}.

\section {Principle of choosing beam energies for long-baseline
neutrino oscillation experiments}

If one wants to resolve the two-fold ambiguity 
($\delta$ $-$ sign of $\Delta m^2_{13}$) in a single experiment 
one cannot tune the experimental parameters so that the $\sin{\delta}$ 
term is maximal because then the trajectory shrinks to a straight line, 
closing the possibility of resolution of the ambiguity. 
Then, there should be a compromise. 
First, one can obtain the conditions to 
maximize lengths of major and minor axes, which lead to \cite{MNjhep01}
\begin{eqnarray}
\left(\frac{E}{\mbox{1 GeV}}\right)_{\cos{\delta}}
 &=& 1.13, 0.47, 0.29
\left(\frac{L}{300\ \mathrm{km}}\right)
\left(\frac{\Delta m_{13}^2}{3 \times 10^{-3}\ \mathrm{eV}^2}\right), \\
\left(\frac{E}{\mbox{1 GeV}}\right)_{\sin{\delta}}
 &=& 0.62, 0.24, 0.14
\left(\frac{L}{300\ \mathrm{km}}\right)
\left(\frac{\Delta m_{13}^2}{3 \times 10^{-3}\ \mathrm{eV}^2}\right).
\end{eqnarray}
It is shown in ref. \cite{MNjhep01} that one can compromize these 
two requirements in the energy region of 0.5 to 2 GeV.

\section {Possibility of simultaneous determination of $\delta$ 
and $\Delta m^2_{13}$ in the super-JHF experiment}

We now address our final subject, probably the most important one, 
the possibility of simultaneous determination of $\delta$ 
and $\Delta m^2_{13}$ in the super-JHF experiment. 
We assume that $\theta_{13}$ is known in a reasonable accuracy 
prior to this experiment.
In fig. 2 plotted is the CP trajectory diagram in 
the number of appearance events plane for 
$\nu_{\mu} \rightarrow \nu_{e}$ and 
$\bar{\nu}_{\mu} \rightarrow \bar{\nu}_{e}$ channels. 
We assume a water Cherenkov detector of  
fiducial volume 0.9 Mton, and 4 MW of proton beam power 
which is planned in the JHF experiment in its phase II \cite {JHF}.
We use the off-axis (OA) 3 degree beam whose neutrino energy peaks 
at $E \sim 0.5$ GeV. 
We refer ref. \cite {MNjhep01} for a detailed explanation of how the 
computation of number of events is done and the results for 
alternative choice of beams.

One can clearly see from fig. 2 that at least 
a half of the parameter space fulfilling the condition 
$\sin{\delta} \cdot \Delta m^2_{13} < 0$
does not suffer from the ambiguity problem. Then, with the 50 \% 
chance (thanks to nature's kind setting!) they will be able to 
determine the CP violating angle 
$\delta$ and the sign of $\Delta m^2_{13}$ simultaneously in 
the super-JHF experiment.

\vglue 0.2cm
\begin{center}
\centerline{\psfig{file=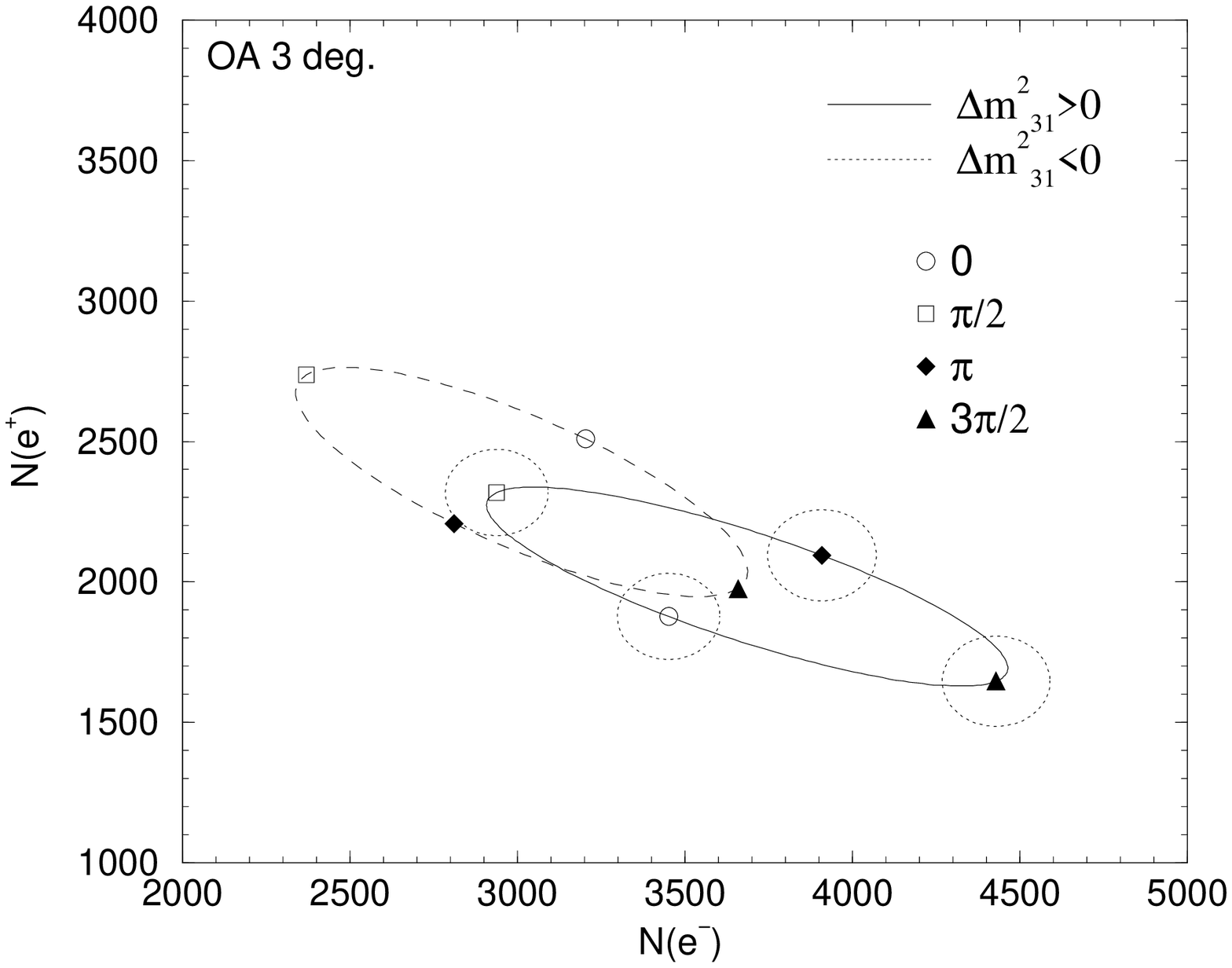,height=12cm,width=12cm}}
\end{center}
\vglue -2.0cm
\noindent{
Fig. 2: CP trajectory in event number plane $N(e^-) - N(e^+)$ 
for OA beam 3 degree and the baseline $L=295$ km. The dotted circles 
correspond to 3 $\sigma$ statistical uncertainty.}
\label{Fig2}


\section*{Acknowledgements}

This work was supported by the Brazilian funding agency
Funda\c{c}\~ao de Amparo \`a Pesquisa do Estado de S\~ao Paulo (FAPESP),
and by the Grant-in-Aid for Scientific Research in Priority Areas
No. 12047222, Japan Ministry of Education, Culture, Sports, Science
and Technology.

\end{document}